\documentclass[epj,twocolumn]{webofc}
\usepackage[varg]{txfonts}   % Web of Conferences font
\woctitle{The 13th International Symposium on Origin of Matter and Evolution of Galaxies}
\begin{document}
\title{An $r-$process macronova/kilonova in GRB 060614: evidence for the merger of a neutron star-black hole binary}

\author{Zhi-Ping, Jin\inst{1}\fnsep\thanks{\email{jin@pmo.ac.cn}} \and
        Yi-Zhong, Fan\inst{1}\fnsep\thanks{\email{yzfan@pmo.ac.cn}} \and
        Da-Ming, Wei\inst{1}\fnsep\thanks{\email{dmwei@pmo.ac.cn}}
        % etc.
}

\institute{Key Laboratory of Dark Matter and Space Astronomy, Purple Mountain Observatory, Chinese Academy of Sciences, Nanjing 210008, China
%\and
%           the second here
%\and
%           Last address
          }

\abstract{%
After the jet break at $t\sim 1.4$ days, the optical afterglow emission of the long-short burst GRB 060614 can be divided into two components. One is the power-law decaying forward shock afterglow emission. The other is an excess of flux in several multi-band photometric observations, which emerges at $\sim$4 days after the burst,  significantly earlier than that observed for a supernova associated with a long-duration GRB. At $t>13.6$ days, the F814W-band flux drops faster than $t^{-3.2}$. Moreover, the spectrum of the excess component is very soft and the luminosity is extremely low. These observed signals are incompatible with those from weak supernovae,
but the ejection of $\sim 0.1~M_\odot$ of $r-$process material from a black
hole-neutron star merger, as recently found in some numerical simulations,
can produce it. If this interpretation is correct, it represents the first time that a multi-epoch/band lightcurve of a Li-Paczynski macronova (also known as kilonova) has been obtained and black hole-neutron star mergers are sites of significant production
of $r-$process elements.
}
%\keywords{gamma-ray burst: individual (GRB 060614) --- radiation mechanisms: thermal ---  binaries: general --- stars: neutron}
%
\maketitle

\section{Introduction}
\label{introduction}
Recently, the Laser Interferometer Gravitational-wave Observatory (LIGO) was replaced with advanced detectors,
the new version known as "Advanced LIGO" has been in operation.
The most promising detections of gravitational waves are expected from collisions and coalescences of neutron stars (NSs) or black holes (BHs).
People paid special attentions on the short-duration (<2s) gamma-ray bursts (GRBs), because they are believed to originate from coalescences of NS binaries or NS$-$BH binaries, and can be precisely localized if their prompt-emission/afterglows are detected.
The evidence for the compact-object merger origin of short GRBs are growing \cite{Berger2014}. Before the successful detection of the gravitational wave radiation, a ``smoking-gun" signature for the compact-binary origin of a GRB would be the detection of the so-called Li-Paczynski macronova (also called a kilonova),
which is a near-infrared/optical transient powered by the radioactive decay of $r-$process material synthesized in the ejecta that is launched during the merger event \cite[e.g.,][]{Li1998,Barnes2013,Tanaka2014}.
Before 2015, just one macronova candidate had been detected in GRB 130603B and such a detection was mainly based on an single epoch observation of Hubble Space Telescope (HST) \cite{Tanvir2013,Berger2013}.

With a duration shorter than 0.2 seconds and without any hint for an extended emission, GRB 130603B is a classical short hard GRB and its association with a macronova is anticipated. However, the situation of GRB 060614 is much more complicated.  It is a nearby burst with a duration of 102 seconds at a redshift $z=0.125$ \cite{Gehrels2006}, and has been classified as a long burst according to its duration. However extensive searches did not find any supernova (SN)-like emission down to limit hundreds of times fainter than SN 1998bw \cite{Galama1998},
the archetypal hypernova that accompanied long GRBs \cite{Fynbo2006,DellaValle2006,Gal-Yam2006}.
Moreover, the temporal lag and peak luminosity of GRB 060614 fall within the short duration
subclass and the properties of the host galaxy distinguish it from other long-duration GRB hosts \cite{Gehrels2006,Gal-Yam2006}. Hence a new classification scheme seems to be needed and it has been called a long-short burst. The physical origin of this type burst has been debated over years \cite{Fynbo2006,Zhang2007}.

Recently, a significant $F814W$-band excess component was reported in a re-analysis of the late time optical afterglow data \cite{Yang2015} of the peculiar event GRB 060614 and has been interpreted as a Li-Paczynski macronova. Like GRB 130603B, such a candidate was mainly based on a single epoch HST observation. In a very recent work by Jin et al. \cite{Jin2015}, the analysis has been improved by considering a possible time evolution of the macronova component and modeling the entire afterglow dataset accordingly. Assuming that only the data in the interval of 1.7- 3.0 days are due solely to the forward shock (FS) emission, and using these data to determine the single power-law decline of the afterglow, the FS component is subtracted from the observational data, a significant excess appeared in multi-wavelength bands at $t>3$ days \cite{Jin2015}. In this proceeding paper we summarize the findings made in \cite{Yang2015,Jin2015} and discuss their implications.

\section{The data analysis result and the excess component}
\begin{table}
\label{tab:macronova}
\begin{center}
\title{}Table 1. The excess component of GRB 060614 (adopted from Jin et al. \cite{Jin2015})\\
\begin{tabular}{lllll} \hline \hline
Time from GRB & Filter & Magnitude$^{a}$\\
(days)    &	& (Vega)  \\ \hline \hline
7.828   & VLT $V$ & (25.6$\pm$0.6)\\
3.869   & VLT $R$ & (25.3$\pm$0.6)\\
4.844   & VLT $R$  & 24.9$\pm$0.3\\
6.741   & VLT $R$ & 25.3$\pm$0.3\\
10.814  & VLT $R$ & (26.5$\pm$0.8)\\
14.773   & VLT $R$ & (27.2$\pm$1.0)\\
3.858 & VLT $I$ & 23.7$\pm$0.4\\
7.841 & VLT $I$ & 24.6$\pm$0.4\\
13.970 & HST F606W ($\sim R$) & 26.9$\pm$0.4 \\
13.571 & HST F814W ($\sim I$) & 25.05$\pm$0.12 \\
\hline
\end{tabular}
\end{center}
Note: a. The magnitudes of the extracted excess component.
The observations with errors larger than 0.5 mag have been bracketed.
\end{table}

The public VLT imaging data of GRB060614 are available from ESO Science Archive
Facility (http://archive.eso.org) while the HST archive data of GRB060614 are available from the Mikulski Archive for Space
Telescopes (MAST; http://archive.stsci.edu) including one visit with WFPC2 and four visits with ACS in F606W
and F814W bands. The details of the analysis of the these raw data can be found in the {\it Methods} of \cite{Yang2015} and the results are summarized in their Supplementary Table 1.

In order to reliably establish the presence of
a distinct HST F814W-band excess in the late afterglow of
GRB 060614, Yang et al. \cite{Yang2015} conservatively assumed that all of the VLT
data were due to the forward shock (FS) and subsequently
fitted all VLT V/R/I-band data at $t > 1.7$ day with the same
decline rate. In such an analysis, only one F814W-band
point at $t\sim 13.6$ days was found to be significantly
in excess of (i.e., the significance is above $3\sigma$) the fitted FS emission. However, the fitted residuals in Fig. 1
of \cite{Yang2015} display an intriguing general trend: the earlier
data ($t < 4$ days) were usually negative (with respect to
the FS afterglow model), while the later data were positive, strongly suggesting that the intrinsic FS emission decline
was likely quicker than that assumed in their model, and
there was very likely to be an excess of emission at $t<13.6$ days. Motivated by such a fact, we have performed an improved analysis of the data.
The main assumption is that only the VLT data in the interval of $1.7-3.0$ days are due to only FS emission, and we used these data to determine the single power-law decline of the afterglow. The physical arguments in support of such an assumption are the following: (i) there was a jet break at $t\approx 1.4$ days \cite{DellaValle2006, Mangano2007, Xu2009},
hence the data collected at later times need be considered; (ii) at $t\sim 1.8$ days a single power-law spectrum can well describe the optical to X-ray data \cite{DellaValle2006,Mangano2007,Xu2009}, suggesting that any macronova contribution to the observed flux is still unimportant; (iii) in the interval of $1.7-3.0$ days there were three measurements in VLT $R$ band and two measurements in VLT $V/I$ bands, with which a reliable estimate of the FS emission decline is plausible. When we subtracted this (best-fitted) FS component from the observational data
we found a significant excess in multi-wavelength bands at $t>3$ days (see Table 1 that is adopted from \cite{Jin2015}). Although the dataset is still relatively sparse, some valuable information can be extracted:
\begin{itemize}
\item The I-band (R-band) flux can reach $\sim 24{\rm th}$ mag ($\sim 25{\rm th}$ mag), corresponding to an extremely-low luminosity $\sim 10^{41}~{\rm erg~s^{-1}}$, and the excess component likely peaked at $t\lesssim 4$ days.
\item The excess component has a very soft spectrum that can be roughly approximated by a thermal one with a temperature $\sim 2700~{\rm K}$.
\item In the time interval of $t\sim 6-13.6$ days the R-band (VLT R and HST F606W) flux dropped with time much faster than that in I-band. For $t>13.6$ days the I-band (HST F814W) flux decline got remarkably steepened and the decline was quicker than $t^{-3.2}$, which clearly did not trace the possible flattening in X-ray band.
\end{itemize}

\section{The excess component: a very weak supernova or a macronova?}
After the discovery of GRB 060614, one widely discussed possibility/mode is that such an event was formed in a new type of a
collapsar which produces an energetic $\gamma$-ray burst that is not
accompanied by a bright SN \cite{Fynbo2006,DellaValle2006,Gal-Yam2006}. The detection of a few very weak SNe (for
example, SN 2008ha \cite{Valenti2009,Foley2009}) with peak bolometric luminosities as low
as $\sim 10^{41}~{\rm erg~s^{-1}}$ is indeed encouraging for such a model.

\begin{figure}
\begin{center}
\includegraphics[width=0.5\textwidth]{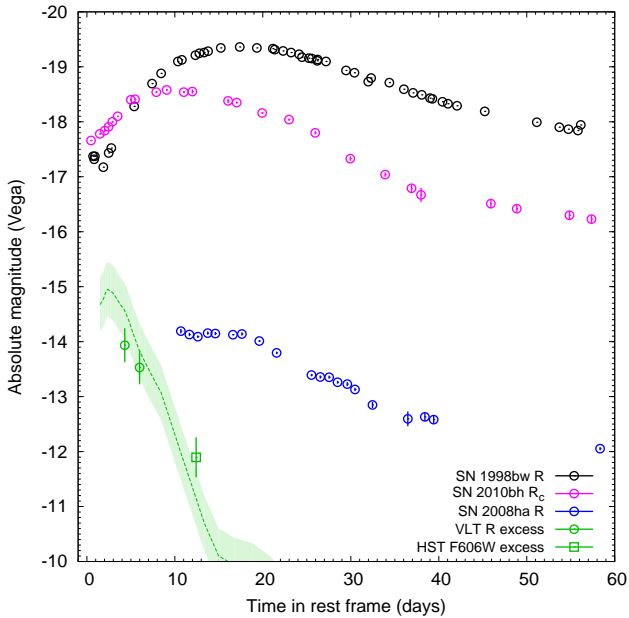}
\end{center}
\caption{The excess component in the afterglow of GRB 060614 compared with some core-collapse SNe.
The data of excess in GRB 060614  are adopted from \cite{Jin2015}. The data of SN 1998bw, the prototype of SN associated with GRB, is taken from \cite{Galama1998};
the data of SN 2010bh, the most rapid to reach the peak SN associated with GRB, is taken from \cite{Cano2011};
and the data of SN 2008ha, the weakest core-collapse SN known so far, is taken from \cite{Foley2009}.
The dashed lines are macronova model lightcurves generated from numerical simulation for the ejecta from a BH$-$NS merger,
with a velocity $\sim 0.2c$ and mass $M_{\rm ej}\sim0.1M_{\odot}$, by \cite{Tanaka2014}, with shadows represent a possible uncertainties of 0.5 magnitudes. The light curve of the excess component is reasonably in agreement with the macronova model, however are quite different to those core-collapse SNe.
} \label{fig:LC}
\end{figure}

Let us compare the excess component of GRB 060614 in R-band (VLT R and HST F606W) with some core-collapse SNe, including some GRB-associated supernovae as well as the extremely weak event SN 2008ha (see Figure \ref{fig:LC}).
The excess emission likely peaked at $t\lesssim 4$ days with an absolute peak magnitude about $R=-13.9\pm0.3$ (see Figure \ref{fig:LC}).
Among GRB-associated SNe, SN 2010bh ($z=0.0591$) had the most rapid rise to maximum brightness, in $R_{\rm c}$-band peaking at $t=8.5 \pm 1.1$ days after GRB 100316D and
the peak magnitude is $R_{\rm c}=-18.60\pm0.08$ \cite{Cano2011,Bufano2012}.
SN 1998bw ($z=0.085$), the prototype of GRB-associated SNe, peaked at $17.3\pm0.2$ days and the absolute magnitude reaches $-19.36\pm0.05$ in $R$-band.
Even compared with SN 2008ha (at a distance of 21.3 Mpc) that had been suggested to be the weakest core-collapse SN \cite{Valenti2009} with absolute magnitude reaches $R=-14.31\pm0.15$ \cite{Foley2009},
the peak magnitude of excess component is just comparable.
Moreover, the R-band decline of the excess component is very fast, which took only about 10 days to decline by 2 magnitudes.
However for SN 1998bw, 2008ha and 2010bh, the decease of the R-band emission by 2 magnitudes took much longer time.

\begin{figure}
\begin{center}
 \includegraphics[width=0.5\textwidth]{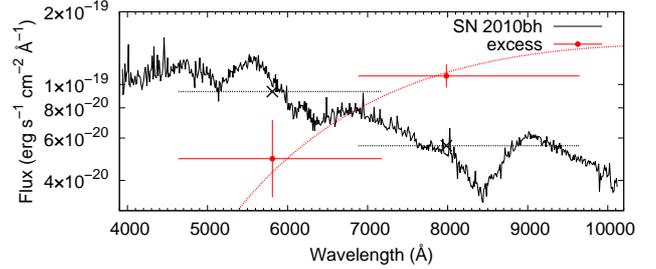}
\end{center}
\caption{The observed SED of the excess component in GRB 060614 compared with the observed spectrum of SN 2010bh.
The excess component in GRB 060614 \cite{Jin2015} and the spectrum of SN 2010bh \cite{Bufano2012} is taken at 12.1 days and 12.7 days in rest frame, respectively.
The spectrum of SN 2010bh has been shifted to $z=0.125$ and rescaled to match the flux of the excess component.
A $T=2700$ K blackbody spectrum is also plotted for comparison (red dotted line),
here the small extinction from the Galaxy (A$_{\rm V}$=0.07) and the host galaxy (SMC like, A$_{\rm V}$=0.05) has been considered to fit for observations.
}
\label{fig:spec}
\end{figure}

In Figure \ref{fig:spec} we compare the spectrum of the excess component in GRB 060614 at $t\sim 13.6$ days with that of SN 2010bh.
Fitted by a blackbody spectrum, the temperature of the excess component is estimated to be $\sim 2700_{+700}^{-500}$ K.
It is much lower than SNe at the same time scale, typically $\sim 5000 -10000$ K\cite{DellaValle2006,Cano2011},
but it is similar to that expected at the photosphere for the recombination of Lanthanides, which is $T\sim2500$ K \cite{Barnes2013}.

All the above facts strongly suggest that the excess component in GRB 060614 is incompatible with being a weak supernova. In the rest of this work we focus on the other possibility, i.e., it is a Li-Paczynski macronova.

\section{Macronova: from the merger of a neutron star binary or a neutron star$-$black hole binary?}
In general a macronova could be powered by either a binary neutron star merger or a neutron star$-$black hole merger.
The major differences of the ejecta from these two type of progenitors are \cite{Hotokezaka2013}:
1) sometimes the NS$-$BH mergers eject much more material than the NS binary mergers; 2) the NS$-$BH merger ejecta is more collimated than the NS binary merger ejecta.
As a result, the macronova powered by an NS$-$BH merger could be much more luminosity and bluer and lasts longer than the macronova powered by an NS binary merger (for instance, see \cite{Tanaka2014}).
In principle it is possible to distinguish between NS$-$BH and NS binary mergers with a multi-band/epoch macronova observation.
As shown in Figure \ref{fig:LC}, a macronova model lightcurve generated from numerical simulation for the ejecta from a BH$-$NS merger,
with a velocity $\sim 0.2c$ and mass $M_{\rm ej}\sim0.1M_{\odot}$, by \cite{Tanaka2014}, can reasonably reproduce the data. To account for the excess component (in particular the F814W-band excess at $t\sim 13.6$ days after the burst) in the NS-NS merger scenario, a $M_{\rm ej}\gtrsim 0.2M_\odot$ is needed. Such a large $M_{\rm ej}$ is well beyond the ejecta mass range found in the NS-NS mergers \cite{Hotokezaka2013} and is thus disfavored.

Therefore, the macronova associated with GRB 060614 is likely from an NS$-$BH merger with very high ejecta mass,
while an NS binary merger seems unable to account for the blue SED as well as the long-duration of the macronova
\cite{Yang2015,Jin2015}. For the macronova signal in GRB 130603B, due to the absence of a multi-epoch/band lightcurve, either a NS-NS merger or a NS-BH merger model can reproduce the data \cite{Hotokezaka2013}.

\section{Discussion}
GRB 060614 was a unique burst straddling both long- and short-duration GRBs and its physical origin was heavily debated over the years \cite{Gehrels2006,Fynbo2006,DellaValle2006,Gal-Yam2006,Zhang2007}.
Recently, a very-soft $F814W$-band excess at $t\sim 13.6$ days after the burst was identified in a joint-analysis of VLT and HST optical afterglow data of GRB~060614 \cite{Yang2015}. It is the {\it first} time to identify a macronova signature in a long-short GRB. An improved analysis of the same set of data reveals a significant excess component emerges at $\sim$4 days after the burst, and present in several late multi-band photometric observations \cite{Jin2015}. Such a signal is clearly incompatible with a very weak supernova (see Figure \ref{fig:LC}). Instead the ejection of $\sim 0.1~M_\odot$ of $r-$process material from a black
hole-neutron star merger, as recently found in some numerical simulations,
can produce it. The binary neutron star merger model is also found to be in tension with the data, either. Therefore we suggest that the excess component identified in \cite{Yang2015,Jin2015} represent the {\it first} multi-epoch/band lightcurve of a Li-Paczynski macronova.

The black hole-neutron star merger interpretation has quite a few far reaching implications \cite{Yang2015,Jin2015}:
\begin{itemize}
\item The presence of
macronovae in both the canonical short burst GRB 130603B and in this `long-short' one, GRB 060614, suggests that the phenomenon is
common and serves as promising electromagnetic
counterparts of gravitational wave triggers in the upcoming
Advanced LIGO/VIRGO/KAGRA era. A conservative estimate of the macronova rate is $\sim 16.3^{+16.3}_{-8.2}~{\rm Gpc^{-3}}{\rm yr^{-1}}$, implying a promising prospect for detecting the gravitational wave radiation from compact object mergers by upcoming advanced detectors, i.e., the rate is ${\cal R}_{\rm GW} \sim 0.5^{+0.5}_{-0.25}(D/200~{\rm Mpc})^{3}~{\rm yr^{-1}}$.
\item A black hole-neutron star merger is favored over the binary neutron star merger in explaining the excess component,
favoring the presence of such binary systems.
\item The so-called `long-short' burst arose from a merger
and not from a collapsar. It was in fact `short' in nature. The fact that a black hole-neutron star merger generates a $\sim 100$ s
long $\gamma-$ray activity is intriguing by itself.
\item The energy source of the Li-Paczynski macronovae is the decay of some unstable $r-$process material. The successful detection of macronovae strongly suggest that such events contribute a significant fraction of
the $r-$process material, which is in support of the model of \cite{Lattimer1974,Eichler1989}.
\end{itemize}

\section*{Acknowledgments}
This work was supported in part by National Basic Research Programme of China (No. 2013CB837000 and No. 2014CB845800),
NSFC under grants 11361140349, 11103084, 11273063, 11433009 and U1231101,
the Foundation for Distinguished Young Scholars of Jiangsu Province, China (Grant No. BK2012047)
and the Strategic Priority Research Program (Grant No. XDB09000000).

%
% BibTeX or Biber users please use (the style is already called in the class, ensure that the "woc.bst" style is in your local directory)
% \bibliography{name or your bibliography database}

\end{document}